\documentstyle[12pt,aaspp4]{article}

\begin{document}

\title{The Interstellar Environment of Filled-Center Supernova Remnants:
	II. G63.7+1.1}

\author{B.J.~Wallace}
\author{T.L. Landecker}
\affil{Dominion Radio Astrophysical Observatory}
\authoraddr{P.~O. Box 248, Penticton, BC, Canada, V2A 6K3}
\and
\author{A.R. Taylor}
\affil{The Department of Physics and Astronomy, The University of Calgary}
\authoraddr{Calgary, AB, Canada, T2N 1N4}

\slugcomment{To appear in the November, 1997 issue of The Astronomical Journal.
No bibliographic reference should be made to this work in pre-print form.}


\date{\today}
\begin{abstract}

A multi-wavelength investigation of the 
candidate supernova remnant G63.7+1.1 and its surrounding
interstellar medium is presented. On the basis of radio continuum data
we conclude that the object is a filled-center supernova remnant, perhaps
in the course of becoming a composite remnant.
The morphology of the remnant, along with HI, $^{12}$CO and high resolution 
IRAS data, suggest that G63.7+1.1 
is interacting directly with the ISM, and does not lie
in a low density region of the ISM. This in turn strongly suggests 
that the detected
nebula is not surrounded by an invisible halo of supernova ejecta.
The association between the SNR and
HI and CO features near the tangent point 
implies a kinematic distance for G63.7+1.1 of $3.8 \pm 1.5$~kpc.

\end{abstract}


\section{Introduction}

Of the 215 Galactic supernova remnants (SNRs) cataloged by Green (1996),
only 9 have been classified as filled-center (FC, also known as ``Crab-like'' 
or ``plerionic''). An object classified as a FC SNR 
must have a centrally brightened radio morphology, a flat
($\alpha>-0.3$), non-thermal, 
radio spectral index, and a complete lack of an associated limb-brightened
shell. It is assumed that the radio emission from these objects is powered 
by a pulsar interior to the nebula, even if no pulsar has been
detected. An object which is centrally brightened but has an associated
limb-brightened shell is classified as a composite remnant. 

The distinction between FC and composite SNRs highlights an important 
problem associated with FC SNRs, the absence of the
limb-brightened shell. If both these types of objects are formed by
supernovae which produce pulsars, 
why should one type have an associated shell but not the 
other? 
The most appealing theory to explain the lack of a
shell around FC SNRs is that put forward by
Chevalier (1977) who proposed that the Crab Nebula (and by extension  
other FC SNRs) consists of two components, a central pulsar-powered core, which
makes up the detected component of the SNR,
and an invisible halo of ejecta, formed at the same time 
as the central pulsar, moving at $\sim 10^4$~km/s. 
He surmised that the fast-moving ejecta around these objects are 
invisible because
the FC SNRs lie in low density regions of the ISM and
the surrounding halos of ejecta have interacted with only minimal amounts of 
material; presumably composite SNRs then lie in more ``normal''
environments and the shell is a direct result.

This hypothesis can be tested by directly imaging the interstellar
medium (ISM) around FC SNRs to determine whether they lie
in low density environments. Romani et al. (1990) used archival IRAS and HI data
and concluded that the Crab Nebula lies in a large-scale, 
low-density void in the ISM. Similarly, Wallace et al. (1994) 
observed the Crab, 3C58, G74.9+1.2 and G21.5-0.9, and found evidence that
all but G21.5-0.9 lie in voids (the data for G21.5-0.9 were inconclusive). 
These studies were
limited by their poor resolution ($\sim 36'$) however, and higher resolution
studies are required to image the ISM distribution immediately around these,
and similar, objects. 

This paper is the second of a series 
in which the ISM around FC SNRs is imaged at high resolution.
The first paper (\cite{wall97}) came to the surprising conclusion that
G74.9+1.2 does {\it not} lie in a low density region of the ISM 
(despite a low density region being suggested by the observations presented in
\cite{wall94}). In this paper we present the results of a 
multi-frequency investigation into the FC SNR candidate G63.7+1.1 and its 
surroundings. G63.7+1.1 was identified as a FC SNR candidate by 
Taylor et al. (1992) based upon its non-thermal radio
spectrum, low ratio of infrared to radio flux density, and morphology (the
entire survey on which Taylor et al. (1992) is based can be found in
\cite{wsrtsurv}).
The data of Taylor et al. (1992) are limited by their relatively poor resolution
however, and further observations are required to confirm the nature of
G63.7+1.1. Presented here are
radio continuum and line, and high resolution far infrared data.
The observations are discussed in Sec.~2 and the data presented in Sec.~3. 
In Sec.~4 the nature of G63.7+1.1 and its surroundings is discussed, and
a model for the interaction between the two is presented. The paper
is summarized in Sec.~5.

\section{Observations}

\subsection{WSRT 1.4~GHz Observations}

As noted above, the observations of Taylor et al. (1992) resolve G63.7+1.1
only poorly (resolution $1' \times 2'$ 
for an object $\sim 8'$ across). To improve our
understanding of this object, G63.7+1.1 was imaged at higher resolution 
using the Westerbork Synthesis Radio Telescope (WSRT) at 1390~MHz.
The observations are summarized in Table~\ref{WSRTobstable}.

The total bandwidth of 80~MHz was split into eight separate bands, each 10~MHz
wide, of which only the central six were usable. The data were edited
and calibrated in the normal fashion at the WSRT and
processed further using the Astronomical Image Processing System (AIPS);
some errors, manifested as radial stripes centered on the
brightest sources, remain in the data. Each 10~MHz band was self-calibrated,
mapped, and CLEANed separately, then averaged to create the final maps.
Images were made of Stokes parameters I, Q and U (Q and U maps were not 
cleaned).

\subsection{IRAS Data}

The infrared data presented here were collected by the Infrared Astronomical
Satellite (IRAS; \cite{IRASexplan88})
and processed using the HiRes procedure
at the Infrared Processing and Analysis Center
(IPAC). The HiRes processing uses an iterative
procedure based on the Maximum Correlation Method (\cite{aumann90})
and can produce images having resolution better than nominal 
for IRAS.  However, the HiRes processing can drastically increase
map artefacts such as striping, which becomes more
prominent as the number of iterations (and the resolution)
increases. The fluxes in the resulting maps are accurate to only
about 25\%, due largely to
difficulty in determining the background level for the region of interest;
this difficulty increases for regions, such as that presented here,
which are close to the Galactic plane. The characteristics of the HiRes IRAS
data presented here are listed in Table~\ref{IRAStable} together with
integrated flux densities of the emission which we associate with G63.7+1.1.

\subsection{CO Observations}

A single observation of the Center for Astrophysics (CfA) 
1.2m CO telescope (\cite{cohen86}; \cite{lt92})
was made towards G63.7+1.1 to look for molecular material possibly associated 
with G63.7+1.1.
The spectrum covers 333~km/s at 0.65~km/s channel separation and the rms
fluctuations of the spectrum are 0.1~K. A sharp peak of strength $\sim
0.65$~K was found at $\sim 11$~km/s, and a broader and weaker peak of strength
$\sim 0.35$~K was found near 23~km/s (all radial velocities are with respect
to the local standard of rest). 

Based on this spectrum,
higher resolution $^{12}$CO observations were made with the Five Colleges
Radio Astronomy Observatory (FCRAO; \cite{erickson92}).  Details of
these observations can be found in Table~\ref{FCRAOtable}.
The background emission was removed by position switching to
a nearby, presumed line-free, position on the sky at
$\alpha = 19^h47^m56^s$, $\delta
= 28^\circ 37' 31''$ (all equatorial coordinates in this paper
are for epoch J2000). 

\subsection{DRAO HI Line Observations}

The HI line data are from observations made at the Dominion Radio Astrophysical
Observatory (DRAO; \cite{roger73}; \cite{veidt85}) in 1987.
Some of the 21 and 74~cm continuum data have been published
by Landecker et al. (1990) while the HI line data are presently unpublished.
Relevant details for the HI data are recorded in Table~\ref{DRAOtable}.
Information on large scale structure, missed by the DRAO synthesis telescope,
is supplied by frequency-switched observations made with the DRAO 26m
radio telescope.
For more information on the observation see Landecker et al. (1990). 

The region of interest falls near the edge of the field. The r.m.s.
noise at map center is 0.9~K, but the primary beam correction raises it
to 3.6~K for the region around G63.7+1.1. 

\section{Results}

\subsection{The WSRT 20~cm Data}

\label{WSRT-data}

G63.7+1.1 (Figure~\ref{fig1}) is approximately $8'$ in diameter
and is centered near $\alpha =
19^h 47^m 55^s$, $\delta = 27^\circ 44'$.
G63.7+1.1 can be thought of as consisting of two components, 
a bright curved ridge running roughly from the south-east to the north-west,
and a fainter diffuse plateau.
The brightness of the plateau drops precipitously around
much of the circumference of the object, but a shallower gradient exists in the 
south-east. The structure of the plateau has a rough bilateral symmetry 
around an axis running from the south-east to the north-west.
A notch in the emission from G63.7+1.1, seen as a slight
bay in the contours immediately to the south of a nearby compact source,
appears in the north of the SNR. A faint tail of emission emerges radially from
the center of the south-east region of the object near a second nearby compact 
source; this tail is aligned with the bilateral symmetry axis of G63.7+1.1. 
While neither compact source can be definitely related to the SNR, it
is interesting to note that the southern source is resolved,
has a position-angle similar to that of the tail, and lies along the 
symmetry axis of the object.

The integrated flux density of G63.7+1.1 is $1.63 \pm 0.06$~Jy at 
1.39~GHz. The quoted uncertainty is strictly internal, and does not
include any possible systematic effects.
A number of flux density values at other frequencies are listed
in Table~\ref{flux.table}.
Earlier observations were made with lower resolution and the two nearby
compact sources are included in those flux determinations.
If those sources are included in our
measurements the total integrated flux 
density is $1.76 \pm 0.01$~Jy, consistent with previous data.

Combining our data with the flux density measurements listed
in Table~\ref{flux.table} we calculate a
spectral index ($\alpha$, $S_\nu \propto \nu^{\alpha}$) of $\alpha =
-0.28\pm0.02$ between 0.408 and 10.55~GHz (systematic effects will probably
result in a larger uncertainty than quoted here) with a possible flattening
at lower frequencies.  
Since the two nearby compact sources
are probably extragalactic, with steep negative spectral
indices, the actual spectral index may be 
marginally flatter than that calculated here. 
Nonetheless, this spectral index demonstrates that the emission 
from G63.7+1.1 is non-thermal in nature (i.e. probably synchrotron 
emission), a conclusion confirmed by the detection of polarized 
emission within the object..

Figure~\ref{fig2} presents vectors corresponding to the 
linearly polarized emission from G63.7+1.1, smoothed to $30''$,
superposed upon contours of the total intensity data.
The average level of polarization across the object is 7\%, reaching a
maximum of 16\% in the north. The level of polarization is lower to the 
south-east
than to the north-west, suggesting greater depolarization of the emission to
the south-east.
The polarization vectors associated with the
brightest emission are oriented perpendicular to the ridge. The vectors 
associated with the plateau region to the north and north-west are
radial, while the vectors in the region in between are circumferential. 
The differences in orientation of these vectors 
suggest either that G63.7+1.1 has an intrinsically complex
magnetic field structure or that there are significant Faraday rotation
differences across the face of the remnant.

\subsection{The IRAS data}

The HiRes IRAS data are presented in Fig.~{\ref{fig3}}.
At 12 and 25~$\mu m$ the emission within the SNR 
is concentrated near the radio peak in the
center of the SNR, while at 60 and 100~$\mu m$ it lies predominately
to the south and east of the peak; the 60 and 100~$\mu m$ emission
appears to form a bowl in
which G63.7+1.1 sits. The emission from these two
areas is not spatially distinct, and independent
fluxes for each region cannot be obtained.
Some compact emission lies to the west of the remnant at 
60 and 100$\mu m$.

The color corrected flux densities for the emission which lies inside
the radio contours are listed in Table~\ref{IRAStable}.
Assuming a $\lambda^{1.5}$ emissivity
law, color temperatures of
$T_{12/25}=183$~K, $T_{25/60}=62$~K, and $T_{60/100}=27$~K are found with
uncertainties on the order of 30\%. A similar wide
variation in the color temperatures is seen in ``old'' SNRs (\cite{sfs92};
\cite{arendt80}), where the typical ``old'' SNR has 
$T_{12/25}=150$~K, $T_{25/60}=60$~K, and $T_{60/100}=30$~K
(younger SNRs tend to have spectra which are reasonably well fit by
a single temperature, $\sim 90$~K). This result
is based on a statistical study of all IRAS-detected SNRs, however, and may
not be strictly applicable to FC SNRs. 

\subsection{The FCRAO CO Data}

The major CO features found in the FCRAO CO data, for the velocity 
range from approximately $-3$~km/s to $24$~km/s, are shown in 
Figure~\ref{fig4}. Three images, chosen to show five CO
features labeled A-E, are presented here.
\begin{description}
	\item[$-$4.7 to $-$2.1~km/s] Feature~A consists of a faint
	arc of CO in the NE portion of the field.
	It is spatially separate from G63.7+1.1 and
	does not appear connected to it in any way.
	\item[ 8.3 to 13.5~km/s] Feature~B is a horseshoe
	shaped feature, centered near
	$\alpha=19^h 48^m 7^s,$ $\delta = 27^\circ 38' 38"$, roughly $8'$ in
	diameter.
	The NW edge of the feature crosses the southern part of G63.7+1.1.
	Interestingly, a gap in the structure
	coincides with the position of the continuum
	``tail'' emanating from the south-east edge of the SNR.
	\item[10.9 to 16.1~km/s] Feature~C appears to the 
		north-west of G63.7+1.1, adjacent to the edge
		of the SNR. It is not clear if this CO is physically
		associated with the G63.7+1.1.
	\item[18.7 to 23.9~km/s] Feature~D appears near the bottom of G63.7+1.1
	roughly coincident with infrared emission at 60 and 100$\mu m$. It
	is interesting to note that Feature~D is coincident in position with
	the gap seen in the north-west of Feature~B.
	\item[21.3 to 26.5~km/s] Feature~E lies in the extreme south-east of
	the observed field.  Once again there does not seem to be any
	evidence of a link between this CO and G63.7+1.1 and there will
	be no further discussion of this feature.
\end{description}

\subsection{The DRAO HI Data}

\label{G63_HI}

After the detection of molecular material possibly associated with
G63.7+1.1, the DRAO HI data cube was searched for indications of atomic
material, in the same velocity range, which might also be associated
with the SNR.  In Fig.~\ref{fig5} the HI data for the region
around G63.7+1.1 for the velocity range 11 to
47~km/s are presented.
The average emission has been subtracted from each channel for presentation
purposes and the data have been smoothed to $2.1'$ circular resolution and
3.3~km/s velocity separation.

While there is no HI emission, above the local background, at the extremes of 
the velocity range shown, emission at the edges of 
G63.7+1.1 is seen in the central velocity maps. 
A plot of the average HI emission over the velocity range
14 to 37~km/s is presented in Figure~\ref{fig6}.
The HI traces the upper boundary of G63.7+1.1 reasonably well. The brightest
HI to the northeast runs parallel to the edge of the SNR, as does much of
the emission to the north-west (although some of the brightest
emission is separated from the remnant, faint emission above
background lies adjacent to the SNR edge); with the present
resolution there is no separation between the outer continuum boundary 
of the SNR and these HI features. The apparent spatial correlation
between the HI emission and the region where the continuum emission
has the steepest intensity gradient suggests
that the emission is physically associated with G63.7+1.1.
HI within the  SNR boundary, seen especially at $\sim 30$~km/s, is adjacent
to CO Feature~D, but differs in velocity by $\sim 10$~km/s..

The average brightness temperature of the HI (above background)
is $T_B = 2.9 \pm 0.7$~K. Assuming a distance to G63.7+1.1 of
$3.8 \pm 1.5$~kpc (see Discussion) the average column density of the
HI, $N_{HI} = 4.4 \pm 1.3 \times 10^{19}$~cm$^{-2}$, implies a mass of
$M_{HI}=51 \pm 43$~M$_\odot$. 
The uncertainty in the mass is 
due mostly to the uncertainty in the distance estimate.

\section{Discussion}

\subsection{The Nature of G63.7+1.1}

With the observations presented here we are able to test the classification
(\cite{twg92}) of G63.7+1.1 as a FC SNR.
An object is generally classified as a FC SNR if it has a flat,
non-thermal, spectral index, a centrally brightened
radio morphology, and no indication of an associated limb-brightened shell.
The spectral index and polarization discussed in Sec.~\ref{WSRT-data}
demonstrate that the emission process is indeed non-thermal.
From Figure~\ref{fig1} it is clear that G63.7+1.1 shows a centrally 
brightened radio morphology (illustrated also in Figure~\ref{fig7}, 
discussed in detail later). There is no evidence of a limb-brightened
shell outside G63.7+1.1 to a level of $\Sigma_{1 GHz} \sim 2 \times 
10^{-22}$~W~m$^{-2}$~Hz$^{-1}$~sr$^{-1}$ (the level of the residual 
uncleaned grating rings), assuming a spectral index of $-0.5$
for any shell emission. A similar limit is calculated from the data of
Taylor et al.~(1996). Thus the minimal criteria for classification
as a FC SNR are met and we interpret the object as a pulsar-powered
nebula (although no pulsar has yet been detected in the SNR).

It is possible, however, to find these same observational 
traits in objects
known as ``interacting composites,'' objects in which a pulsar has
caught up to, and is interacting with, the SNR shell created in the
same explosion as the pulsar (\cite{sfs89}). 
This class of objects is of necessity
associated with fast moving pulsars, and the resulting nebulae reflect this
with a distinct bow shock morphology 
oriented along the direction of motion of
the pulsar. In addition, there is usually an indication of the existence of
the SNR shell, either in the form of ``wings'' of emission associated
with the FC nebula (e.g. CTB80, \cite{s-h95}), 
or as a fully formed shell
located near the FC nebula (e.g. G5.4-1.2; \cite{fk91}). 
G63.7+1.1 shows no indication of either of these phenomena. While part of 
G63.7+1.1 has distinctly sharper edges than the rest, the characteristic
``v''-shape of a bow-shock is not present. Similarly, there is no indication
of an associated SNR shell in either the 20cm data presented here, nor in
the wider field 92cm data of Taylor et al. (1996). Consequently, we conclude 
that G63.7+1.1 is not an interacting composite SNR.

\subsection{The Morphology of G63.7+1.1}

The steep intensity
gradient  at the edge of the northern section of the remnant,
compared to the shallow gradient to the south-east,
suggests an interaction between the SNR and its surrounding ISM.
To better illustrate the nature of the sharp intensity gradient, the SNR was
divided into 12 sectors, each $30^\circ$ wide, and
an average radial profile for each sector was derived. 
Figure~\ref{fig7} shows two average intensity profiles through 
the SNR. The shallower profile (Fig.~\ref{fig7}a)
is centered to the south (at 6 o'clock), while the steeper
profile (Fig.~\ref{fig7}b), 
typical of the profiles taken from the northern portion
of the SNR, is centered to the west (at 3 o'clock).
The southern profile declines smoothly to zero and is essentially
featureless while the western profile reaches a plateau at $\sim 4$~mJy/beam
before the
brightness drops to zero. The region over which the
brightness drops from 4~mJy to zero in Fig.~\ref{fig7}b 
is spatially unresolved.
The difference in distance, from assumed SNR
center, at which the two profiles reach zero reflects the asymmetric 
morphology of the SNR.

Assuming optically thin emission, the brightness distribution within the
remnant is a function of 
the volume emissivity distribution and the observed path-length 
through the SNR. As the path-length 
presumably decreases towards the edge of the SNR,
a uniform or radially decreasing emissivity should result in the brightness
decreasing towards the edge. The
constant brightness across the plateau implies an
emissivity which increases with radius, an interpretation that 
is at odds with the commonly accepted view of FC SNRs as
homogeneous bubbles of synchrotron emitting material (e.g. \cite{ps73};
\cite{rc84}).

The emissivity was explored quantitatively 
by deconvolving the 2-D brightness distribution
with the assumed 3-D shape of the SNR (\cite{lv94}; \cite{leahy91}) 
using software kindly provided by D.A.~Leahy.
The emissivity profiles resulting from the
intensity profiles of Figure~\ref{fig7} are presented
in Figure~\ref{fig8}. The emissivity profile for the
southern sector (Fig.~\ref{fig8}a)
shows a peak at or near the center of the SNR,
followed by a slow decline towards the edge of the SNR. The 
profile for the western wedge (Fig.~\ref{fig8}b)
also shows the central emissivity peak, but
with a secondary emissivity peak near the edge of the SNR. Most of the
other emissivity profiles (not shown) also exhibit this peak in
emissivity towards the edge of the SNR, although to lesser extents. 

The rise in emissivity towards the edge of the SNR can be explained
in two ways, both of which imply an interaction between the SNR and
its environment. Assuming a particle energy
distribution of the form $N(E)=N_\circ E^{-\gamma}$ 
(where $E$ is the electron energy, $N(E)$
the number of electrons with that energy, and $N_\circ$ and $\gamma$
constants), the volume emission coefficient 
($\epsilon_\nu$) is given by
\begin{equation}\epsilon_\nu = f(\gamma) N_\circ \nu^{1-\gamma/2} 
(B \sin{\theta})^{\gamma +1/2}\end{equation}
(\cite{pacho}) where 
$\theta$ is the pitch angle of the electrons, $B$ the magnetic field intensity,
and $f(\gamma)$ is a complicated function of $\gamma$.
An increase in either the magnetic field strength or the particle
energy distribution (by modifying either $N_\circ$ or $\gamma$) 
can then result in an increase in the volume emissivity. 

The magnetic field strength can be increased 
if the SNR is interacting with the ISM and compressing the ambient magnetic 
field. Additionally, an interaction between the SNR and the ISM may result
in the creation of a second population of relativistic 
particles at the edge of the SNR. If G63.7+1.1 is
similar to the Crab Nebula, the detected nebula is bounded by a thin
shell of slow-moving ($10^3$~km/s)
SN ejecta which drives a blast wave into
the surrounding ISM. The interaction between the blast wave and the
ISM may produce relativistic particles near the
shock as for a shell SNR. This would result in increased numbers of 
emitting particles 
at the outer boundary of the synchrotron nebula, as compared to the interior
of the nebula, thus increasing the emissivity as required. 

This second scenario raises an interesting possibility.
The power index of the particle energy distribution, $\gamma$,
is different for particles produced by the pulsar and for those 
produced by the interaction
with the ISM (for the pulsar accelerated particles $\gamma \sim 1.4$, for
the shock accelerated particles, $\gamma \sim 2$). 
This second population of particles should then result in an
observable spectral steepening at the edge of G63.7+1.1, similar to
that found in the composite SNR G18.95-1.1 (\cite{furst97}).
If such steepening exists then G63.7+1.1 will, at some frequency, show
limb-brightening, suggesting that G63.7+1.1 may be undergoing a
transition from FC to composite SNR, providing the ``missing link'' between 
these two classes of objects. 

We interpret the rise in emissivity as a result of the interaction
between the synchrotron nebula and the ISM, and {\it not} as a result of
an interaction between a fast moving halo of ejecta and the ISM. 
A blast wave associated with fast moving ejecta would
have disrupted the surrounding material and swept it away from the synchrotron
nebula. While recombined HI has been found behind the shock in W44 
(\cite{kh95}) the blast wave which ionized the material is
easily visible; this is clearly not the case with G63.7+1.1, supporting our
interpretation.

\subsection{A Model for G63.7+1.1}

Our information on the ISM comes from three sources: the FIR, CO and 
HI data, probing the dust, molecular and atomic components respectively.
Much of the emission in these data is spatially coincident with 
G63.7+1.1. While it can be argued that these coincidences are the result of
chance superposition of unrelated features along the line of sight to the
SNR, several aspects of the data argue for a physical connection between the
SNR and its surrounding ISM.

To illustrate this connection Fig.~\ref{fig9} shows contours of
the emission from the radio continuum, 100$\mu m$, and CO feature~D. 
The FIR emission provides the strongest argument for an association. At 60
and 100$\mu m$ the emission forms a bowl around the bottom of the G63.7+1.1,
with the northern edge of the emitting region lying adjacent to the southern
extreme of the bright radio continuum ridge.
At 100$\mu m$ a finger of emission extending to the northwest is seen near 
$\alpha \sim 19^h 45^m$, $\delta \sim 27^\circ 43'$; 
this finger lies adjacent to the bright ridge, and extends in the same
direction (Fig~\ref{fig9}). A similar finger is seen in CO feature~D.
Combined with the color temperatures of the dust emission,
which are consistent with the shocked dust
associated with other SNRs, these characteristics suggest a causal 
relationship between the FIR emitting material, CO Feature~D, and the SNR.

These aspects of the data provide compelling
arguments for an association between G63.7+1.1 and molecular material to the
south. The HI data provide additional, albeit weaker, support for
this association. The agreement in morphology between the partial HI shell 
seen in Fig.~\ref{fig6} and G63.7+1.1 argues for a relationship between
the two. In addition, the faint HI emission seen interior to the SNR at
$v \sim 30$~km/s lies adjacent to CO Feature~D; this feature could be unrelated,
or it could be molecular material from Feature~D which has been dissociated
and ionized by the interaction with G63.7+1.1, and which has then recombined.
While this would explain the velocity shift between the HI and CO emission, the
small size of this velocity shift suggests that the material has been
accelerated only a small amount, implying that the shock associated with the
interaction was weak. In turn, this suggests that any spectral steepening
associated with the SNR/ISM interaction may be small.
The large noise in the HI data do not allow us to say
anything more about this possibility.

In short, we feel that these observations show that G63.7+1.1 is
indeed interacting with its surroundings and allow us to present
the following observational scenario.

The SNR G63.7+1.1 lies adjacent to molecular material to
the south (Feature D at 21~km/s) 
with which it has interacted. The SNR shock has heated some of the
dust associated with the molecular material, giving rise to the FIR emission.
The shock has dissociated and ionized
some fraction of the molecular material, accelerating
it in the process by $v\sim 10$~km/s, and the recombined material gives
rise to the faint HI emission seen interior to the SNR. The
HI to the north of the remnant at these same velocities mimics the
morphology of the SNR, possibly the result of the interaction between 
the SNR and
the surrounding atomic material. The expansion of the SNR is truncated 
to the south-west
by its interaction with the molecular material, while the HI forms
a ``helmet'' surrounding the SNR in all other directions. The
projection of the FIR and CO emission within the SNR suggests that the
whole system is inclined at an angle to the plane of the sky, resulting
in the observed morphology.	
The situation is similar to that described for another
FC SNR, G74.9+1.2 (\cite{wall97}).
That SNR lies near an HI shell which defines the inner edge of an interstellar
cavity; on one side the pulsar powered nebula expands freely into the low
density inside the cavity, but on the other side the HI shell
impedes the expansion.

This model allows us to calculate a distance to G63.7+1.1 based upon the
radial velocity of the associated ISM. The central velocity of both the
HI ($v \sim 25$~km/s) and CO Feature~D ($v \sim 21$~km/s) are close
to the velocity of the tangent point in this direction ($v=22.4$~km/s).
The velocity ranges quoted for both the
CO and HI data extend beyond this velocity, suggesting that G63.7+1.1
lies at or near the tangent point. Assuming a flat rotation curve 
(R$_\circ = 8.5$~kpc and V$_\circ =220$~km/s) the distance to
the tangent point is $3.8 \pm 1.5$~kpc; the velocity at which CO Feature~D is
first seen (18.7~km/s) corresponds to a kinematic distance of either
2.3 or 5.3~kpc, for the near and far sides of the tangent point distance
respectively, leading to the large uncertainty.
The linear diameter of G63.7+1.1 is then $8.8$~pc
and the luminosity $4.3 \times 10^{-21}$~W~Hz$^{-1}$.

Situations such as we propose for G63.7+1.1 are
modeled by Arthur \& Falle (1991, 1993) who
suggest that the remnant of a SN which explodes near a density 
discontinuity, such as the edge of a molecular cloud,
will be roughly elliptical, with the SNR penetrating further into
the low density region than into the high density region. The
symmetry axis of the remnant will be 
perpendicular to the plane of the density interface.
These aspects of the model agree with the observed morphology of the
SNR and with its location relative to CO feature~D.

The models of  Arthur and Falle predict the
presence of a jet of material, ``squirted'' from the high density region
into the low, near the symmetry axis of the SNR.
Such an effect may be responsible 
for the observed bright ridge along the symmetry axis of G63.7+1.1.
If molecular material
has been squirted into the interior of the SNR, as suggested by
Fig~\ref{fig9}, then some portion of
it may be expected to
be dissociated and ionized, providing additional particles which can
contribute to the radio continuum emission to form the observed ridge.
If this is the case then emission associated with the ridge may
have higher Faraday rotation than the rest of the remnant due to the
presence of a large amount of ionized material.
Radio emission from the south-eastern portion of the SNR may 
be depolarized by strong Faraday rotation associated with 
dense ionized material created
as the SNR interacts with the molecular material to the south; it is 
interesting to note that the 60$\mu m$ FIR emission appears to delineate the
region of strongest depolarization. This depolarization
would imply that the molecular material must be on the near side of the
SNR, and that the SNR is inclined away from the observer. 
Measurements of the intrinsic magnetic field
orientation and the rotation measure of the SNR based on
radio polarimetry at short wavelengths may help
cast light on these issues. 

The tail of emission to the south-east may be
similar to the Crab Nebula's chimney, and may arise from the
same process (e.g. \cite{cox91}). However,
if the SNR is interacting with CO Feature~B (instead of, or in addition
to, Feature~D, as
assumed elsewhere in the paper) an additional possibility arises.
It has been noted that the ``tail'' coincides in position
with a ``gap'' in the CO feature. The tail may be the result of
SN ejecta being impeded by the molecular cloud but escaping through
the gap. 

\subsection{Implications}

Two pieces of evidence suggest strongly that G63.7+1.1 is interacting with 
the surrounding ISM. First, there is the steep intensity gradient around much 
of the perimeter of the SNR. Second, there is the striking emission at
60 and 100$\mu m$ which bears a strong morphological resemblance to the SNR;
the fingers of emission seen in the 100$\mu m$ data and in CO Feature~D, 
both of which extend in the same direction as the radio continuum ridge, 
add further support to this suggestion.
The apparent interaction between G63.7+1.1 and the surrounding interstellar 
material illustrates that this FC SNR is {\it not} expanding into
a low density region of the ISM. This makes G63.7+1.1 the second FC SNR
for which this can be said (the first is G74.9+1.1; \cite{wall97})
and has important implications for our understanding of the nature of FC SNRs.

Chevalier (1977) invoked low density ISM to explain the absence of 
shells associated with FC SNRs. The assumption made is that no
emission will be detectable if the blast wave
associated with a fast moving halo of ejecta has encountered only a small
amount of material. However, it has been shown that G63.7+1.1 and G74.9+1.2
do not lie in low density regions of the ISM. This implies directly
that both these objects are not surrounded by fast moving halos of ejecta, and
thus that Chevalier's (1977) model is not applicable for them (although it
may still hold for other FC SNRs). Since
most models of FC SNRs (e.g. \cite{rc84};
\cite{kc84}) assume that they are surrounded
by fast moving ejecta, a refutation of Chevalier's
hypothesis demands a rethinking of even our most basic
understanding of these objects.

\section{Conclusion}

High resolution radio continuum observations of the SNR candidate G63.7+1.1
have been presented, along with far-infrared, HI, and CO observations of its
surroundings. These observations show that G63.7+1.1 has
a centrally-brightened radio morphology with no detectable limb brightening;
no limb brightened shell is seen down to a level of 
$\Sigma_{1\ GHz} \sim 2 \times 10^{-22}$W~m$^{-2}$~Hz$^{-1}$~sr$^{-1}$.
The spectral index of the emission, $\alpha=-0.28 \pm 0.02$, along with
the detection of polarization, imply a non-thermal emission process. 
G63.7+1.1 is thus most likely a filled-center SNR.

The morphology of G63.7+1.1 can be broken down into two components, a
bright resolved core and a fainter, more uniform, plateau. The plateau is
shown to have a sharp, unresolved, intensity gradient marking its outer 
edge around most of
the remnant, but exhibits a much shallower gradient to the south-east. 
Modeling of the emissivity profile within the remnant suggests that
an increase in emissivity must occur around the portions of the remnant having
a sharp intensity gradient. This rise in emissivity implies an increase
in magnetic field strength, the injection of additional emitting particles,
or both, at the edges of the SNR. This may suggest that G63.7+1.1 is 
in the process of forming a limb-brightened component at its edges and may
thus be in transition from FC to composite SNR.

The observations of the ISM around G63.7+1.1 suggest that the remnant lies
adjacent to molecular material with which it is interacting. Some of
this molecular material may have been injected into the SNR interior, 
perhaps creating a ridge of brighter radio emission with associated
features seen at 100$\mu m$ and in CO~Feature~D. The radial velocity of the
molecular material with which the SNR is interacting
indicates that the SNR lies at or
near the tangent point of the Galactic rotation curve in this direction,
implying a distance to the remnant of $3.8 \pm 1.5$~kpc.

The interaction between G63.7+1.1 and the ISM implies both that the SNR
is {\it not} lying in a low density region of the ISM, and that the SNR
is {\it not} surrounded by a fast moving halo of ejecta. This result demands 
a rethinking of models used to describe FC SNRs and their evolution.

\acknowledgments
The authors would like to thank Daniel Puche for obtaining the CfA
spectrum towards G63.7+1.1. Gerald Moriarity-Schieven is acknowledged for
his invaluable assistance in obtaining the FCRAO data. Grateful thanks
are offered to Mark Heyer for obtaining additional
FCRAO data for the authors. 
Thanks are also due to Denis Leahy for provision of the software used in
Sec.~4, and to Wolfgang Reich for providing his 4.85 and 10.55~GHz
flux densities. The Dominion Radio Astrophysical Observatory
is operated as a national facility by the National Research Council of Canada.
This research was supported in part by a grant from the Natural Sciences
and Engineering Research Council of Canada.

\vfill\eject

\vfill\eject

\figcaption[Wallace.fig1.ps]{The large extended feature in the middle of the
       map is G63.7+1.1. The contours are at -0.2 (dashed),
       0.2, 0.6, 1.0~mJy/beam
       and upwards in steps of 1~mJy/beam thereafter.
       Note the two compact sources to the north
       and south-east of the remnant as well as the artefacts
       emanating radially
       from the bright compact source to the south-west.
\label{fig1}}

\figcaption[Wallace.fig2.ps]{The lines denoting polarized electric
		field intensity
		and position angle are superposed upon the
		contours from the total intensity data. The polarization
		data has been smoothed to $30''$ circular resolution while 
		the total intensity contours are at full resolution. The
		position angle of the lines denotes the direction of
		the electric field vectors, and a vector of length $1'$ 
		corresponds to a linearly polarized intensity of
		approximately 1~mJy/beam.
\label{fig2}}

\figcaption[Wallace.fig3.ps]{The upper panels show the 12 and 25$\mu$ data
	in greyscale (left and right respectively) while the lower
	show the 60 and 100$\mu$. Contours from the WSRT 1390~MHz
	continuum image are overlaid for reference. 
\label{fig3}}

\figcaption[Wallace.fig4.ps]{The FCRAO CO data are presented in greyscale,
        with contours of the WSRT continuum image
        overlaid for reference. The top figure is the average of two
	channels while the bottom two figures are three channel averages
	centered on the velocities indicated. The greyscale runs from -0.3 to
	+0.7~K, the contours run from 1 to 19~mJy/beam in steps
	of 2~mJy/beam. The features labeled in the plot are discussed in the
	text.
\label{fig4}}

\figcaption[Wallace.fig5.ps]{The greyscale, denoting HI emission, ranges from
		30~K below the background (white) to 25~K above (black).
		Overlaid are
		contours of the WSRT 1390~MHz continuum image. The central
		velocity of each image is indicated.
\label{fig5}}

\figcaption[Wallace.fig6.ps]{The HI emission (greyscale) averaged over
	velocities 14 to 37~km/s; overlaid are contours from the
	WSRT 21cm data.  The most intense HI emission lies around
	the northern edge of the SNR.
\label{fig6}}

\figcaption[Wallace.fig7.ps]{Two intensity profiles through G63.7+1.1. Profile A
		is from a sector centered on the southern
		portion of the remnant, while B is from
		the western portion. Notice that the intensity reaches a
		plateau in B at $\sim 4$~mJy/beam. 
\label{fig7}}

\figcaption[Wallace.fig8.ps]{The emissivity profiles derived from the
	intensity profiles presented in the previous figures. Notice
	that the secondary emissivity peak at $\sim 3'$ in B is not seen in A.
	Arbitrary units are used for the emissivity scale.
\label{fig8}}

\figcaption[Wallace.fig9.ps]{Contour plots of the radio and 100$\mu m$ emission,
	along with that associated with CO Feature~D. The line extending from
	south-east to north-west was chosen to match the finger of
	100$\mu m$ emission discussed in the text. The perpendicular line
	was chosen to pass through the peak of the 100$\mu m$ emission.
	Features parallel to the 100$\mu m$ finger can be seen in the
	radio continuum and in CO Feature~D.
\label{fig9}}

\vfil\eject

\tablenum{1}
\tablewidth{0pt}
\begin{deluxetable}{l c}
\tablecaption{Summary of WSRT 1390~MHz Observations Towards G63.7+1.1}
\startdata
Field Center & $\alpha(J2000) = 19^h 45^m 55.0^s$  \\
 & $\delta(J2000) = 27^\circ 36' 30.0"$ \\
Bandwidth (observed/used) & 80/60 MHz  \\
Resolution (RA$\times$Dec)& $13.8'' \times 26.4''$  \\
RMS deviations & 50~$\mu$Jy/beam  \\
Shortest Baseline & 36m  \\
Largest Detectable Structure & $\sim 20'$  \\
\enddata
\label{WSRTobstable}
\end{deluxetable}

\tablenum{2}
\tablewidth{0pt}
\begin{deluxetable}{l c c c c}
\tablecaption{IRAS Data for G63.7+1.1}
\startdata
 & $12\mu$ & $25\mu$ & $60\mu$ & $100\mu$ \\
Resolution (RA $\times$ Dec) & $2.5' \times 1.0'$ & $1.6' \times 0.5'$ &
$2.75' \times 1.1'$ & $2.8' \times 2'$ \\
Iterations & 1 & 5 & 5 & 20 \\
Color Corrected Fluxes (Jy) &$5.7 \pm 1.0$ & $6.3 \pm 1.2$  & $28.8 \pm 2.5$ &
$105 \pm 33$ \\
\enddata
\label{IRAStable}
\end{deluxetable}

\tablenum{3}
\tablewidth{0pt}
\begin{deluxetable}{l c c}
\tablecaption{Summary of FCRAO $^{12}$CO Observations Towards G63.7+1.1}
\startdata
Field Center & RA(J2000) & $19^h 47' 57.5"$ \\
 & Dec(J2000) & $27^\circ 44' 0.8''$ \\
Field Size & RA $\times$ Dec & $20.'9 \times 20.0'$ \\
Resolution & & $50.2''$ \\
Beam Spacing & & $25.1''$ \\
Central Velocity & & $20$~km/s \\
Number of Channels & & 32 \\
Channel Separation & & 2.60~km/s \\
Integration time & & 60s \\
per point & & \\
\enddata
\label{FCRAOtable}
\end{deluxetable}

\tablenum{4}
\tablewidth{0pt}
\begin{deluxetable}{l c c}
\tablecaption{DRAO HI Observations Towards G63.7+1.1}
\startdata
Field Center & RA(J2000) & $19^h 46^m 18.3^s$ \\
 & Dec(J2000) & $28^\circ 42' 23.4''$ \\
Synthesized Beam  & RA $\times$ Dec & $1.'0 \times 2.'1$ \\
Smoothed Resolution & RA $\times$ Dec & $2.'1 \times 2.'1$ \\
Central Velocity & & $-24.7$~km/s \\
Channel Separation & & 1.65~km/s \\
Polarization & & RHCP \\
Observation date & & 1987 \\
\enddata
\label{DRAOtable}
\end{deluxetable}

\tablenum{5}
\tablewidth{0pt}
\begin{deluxetable}{c c c}
\tablecaption{Measured Radio Flux Densities for G63.7+1.1}
\tablehead{
\colhead{Frequency} & \colhead{Flux Density} & \colhead{Uncertainty} }
\startdata
327 \tablenotemark{a} & 1.47 & 0.04 \\
408 \tablenotemark{b}& 2.1 & 0.2  \\
1390 \tablenotemark{c}& 1.63 & 0.06 \\
1420 \tablenotemark{b}& 1.71 & 0.02 \\
2695 \tablenotemark{d}& 1.41 & 0.1  \\
4850 \tablenotemark{e}& 1.16 & 0.10  \\
10550 \tablenotemark{e}& 0.95 & 0.05 \\
\enddata
\tablenotetext{a}{Value obtained by the authors from the data of Taylor et al.
                1996}
\tablenotetext{b}{Value obtained by the authors from unpublished data of
                Landecker et al. 1990}
\tablenotetext{c}{This paper}
\tablenotetext{d}{Value obtained by the authors from the data of Reich et al.
                1990}
\tablenotetext{e}{W. Reich (private communication)}
\label{flux.table}
\end{deluxetable}

\vfill\eject

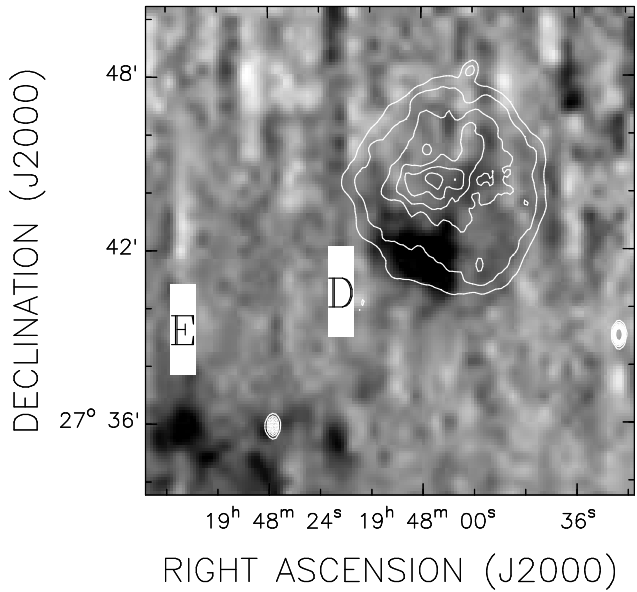
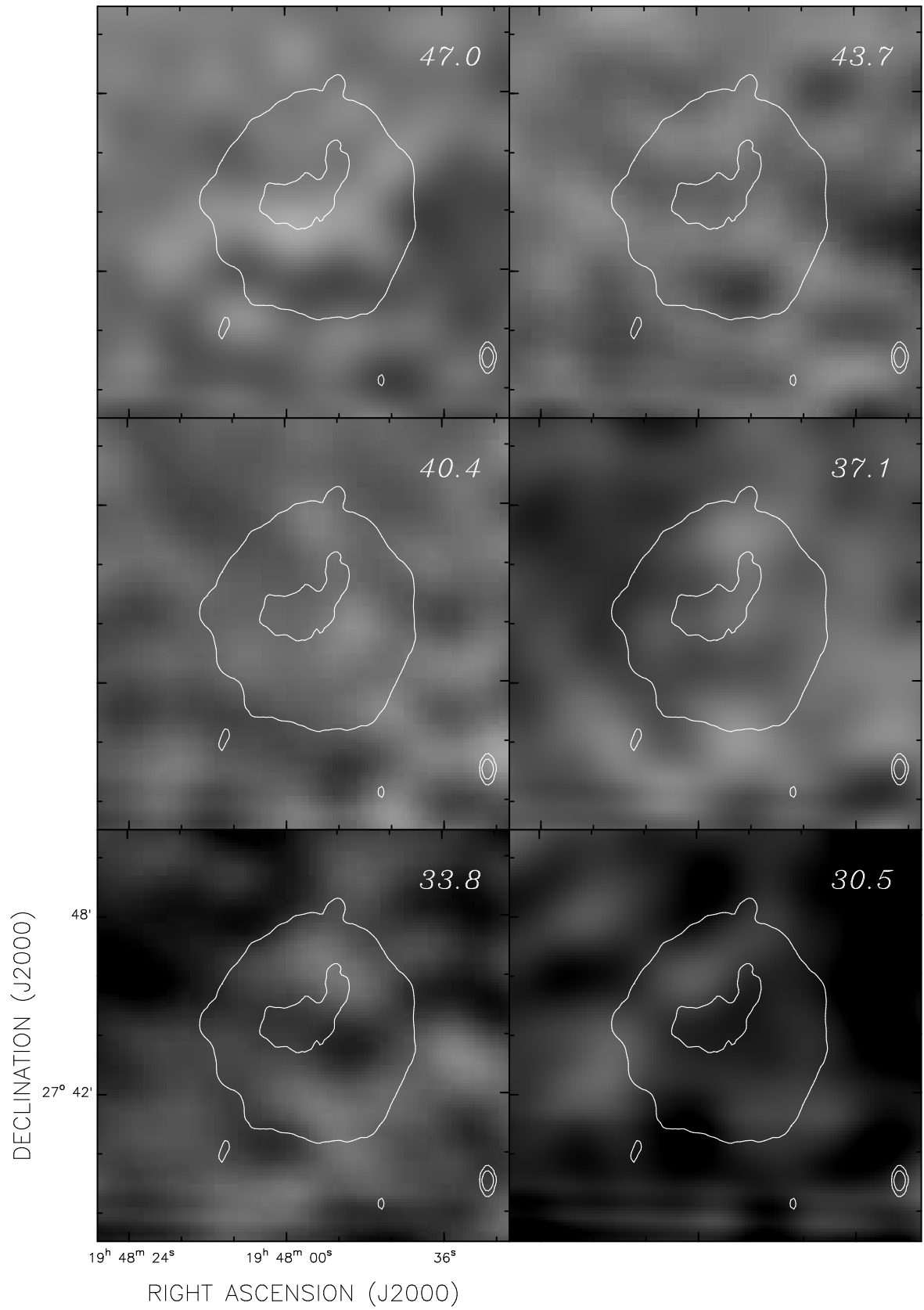
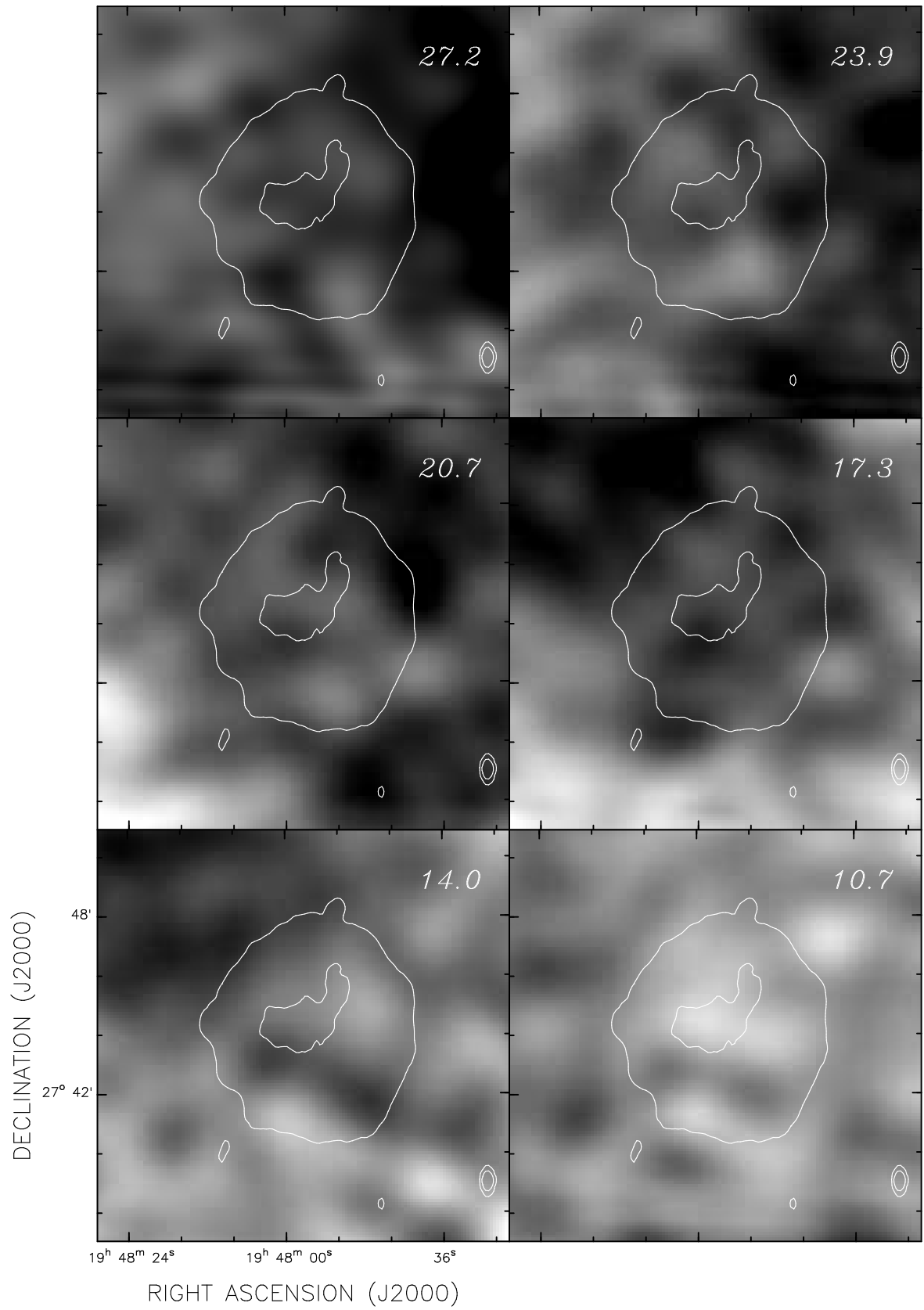
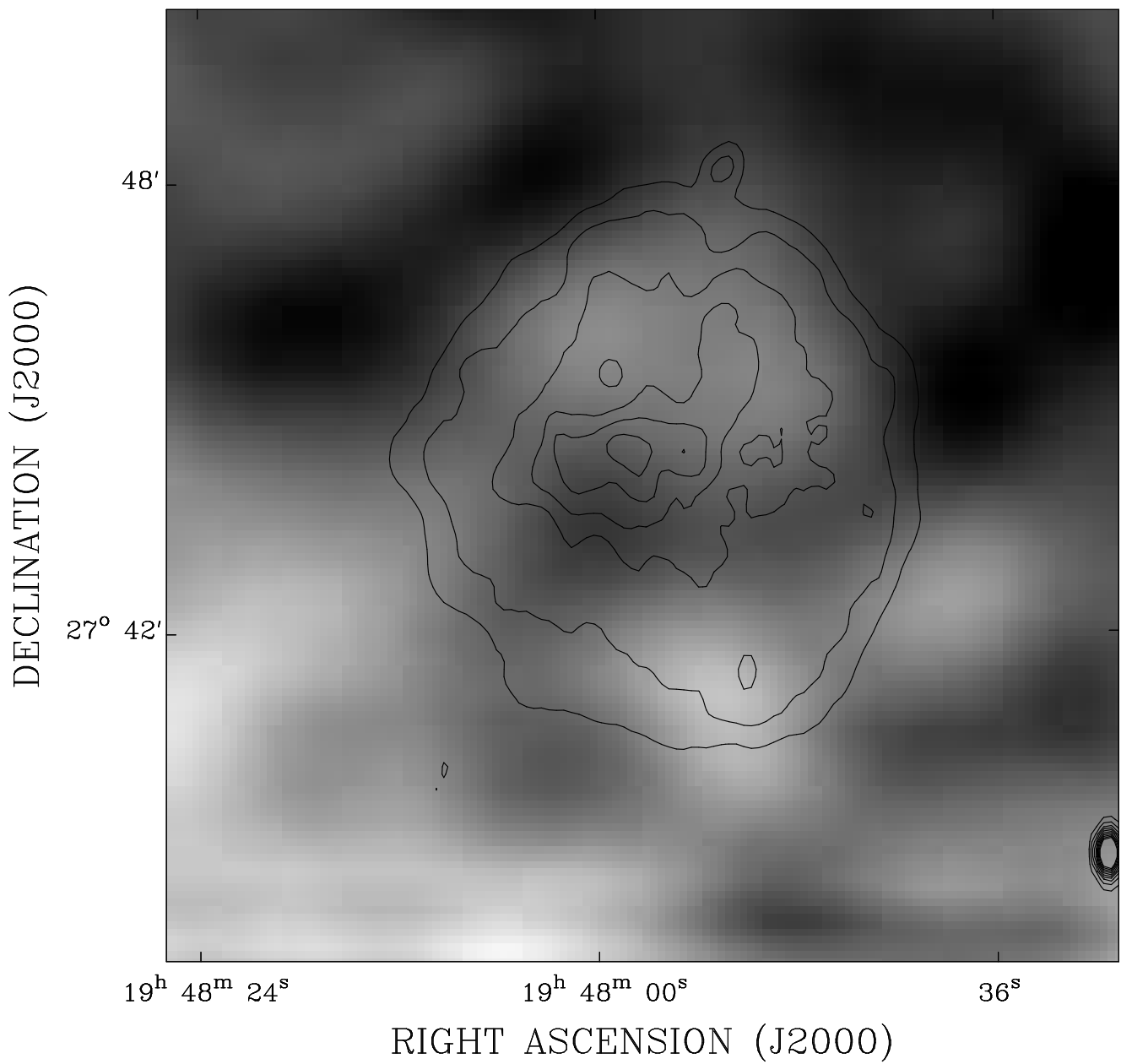
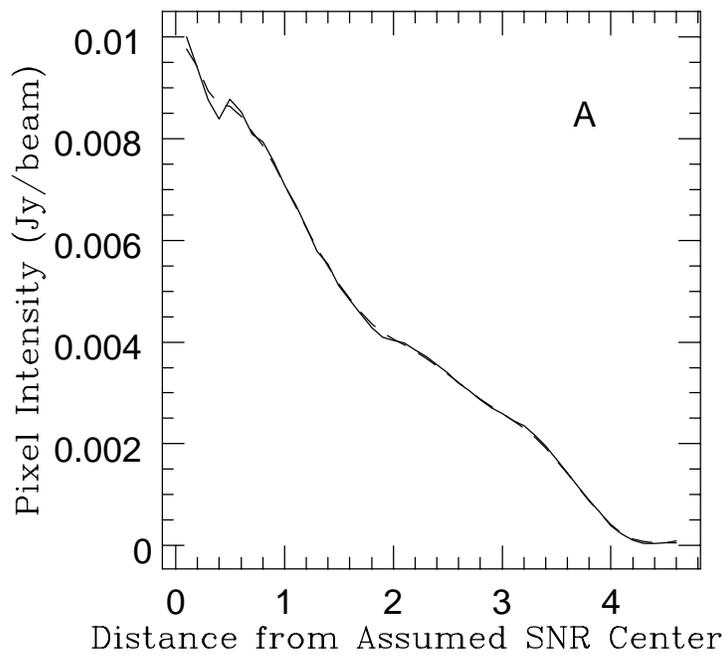
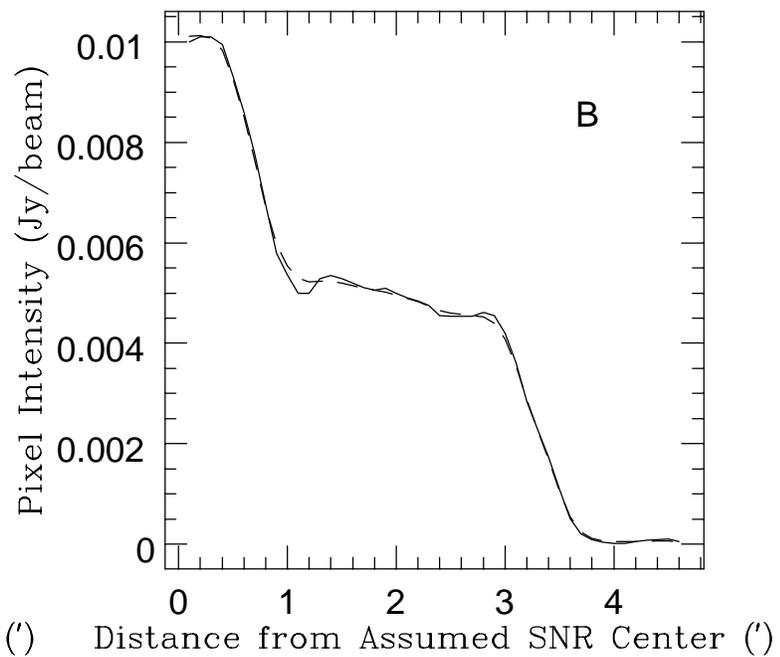
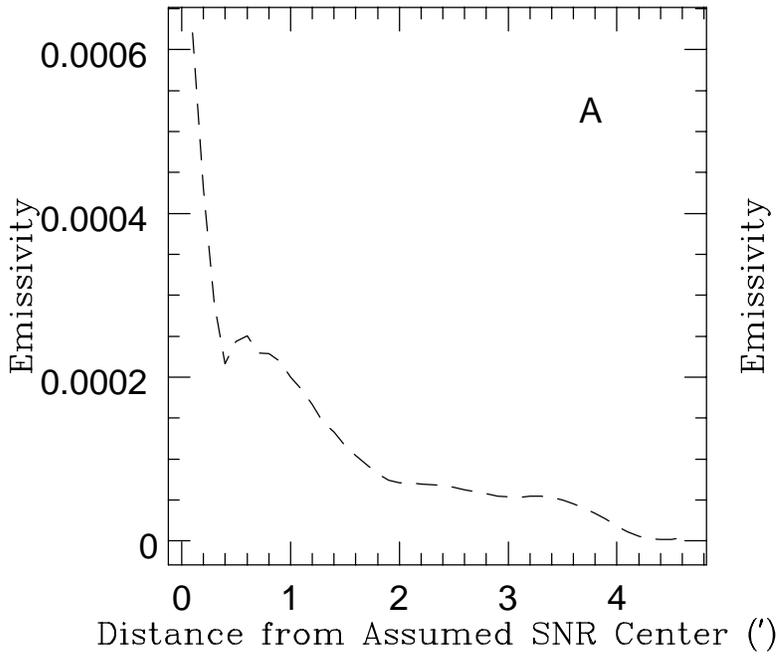
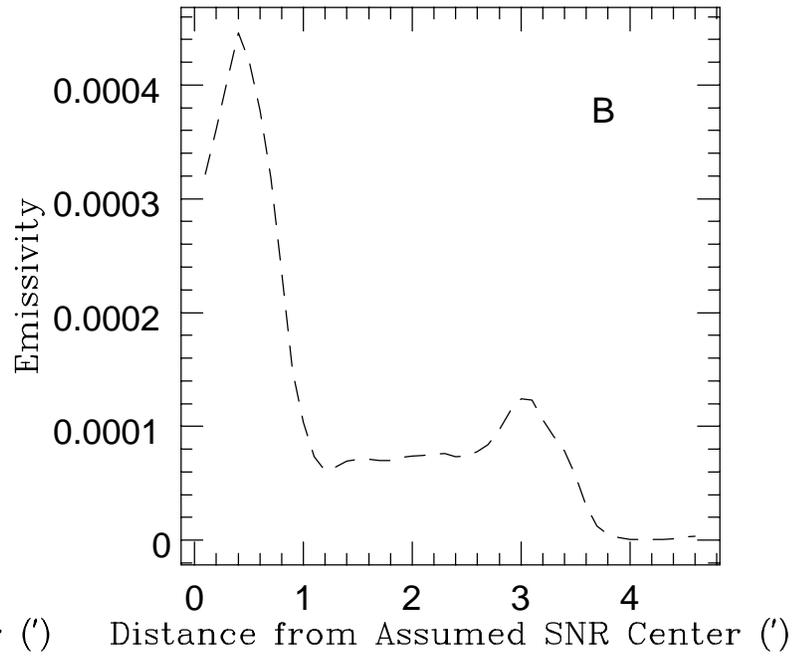
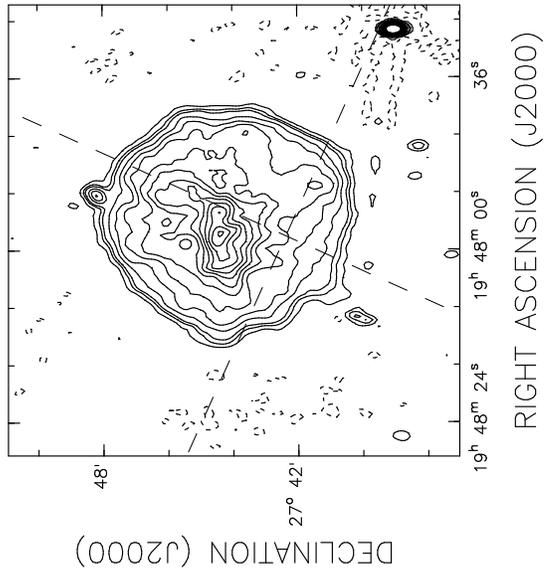
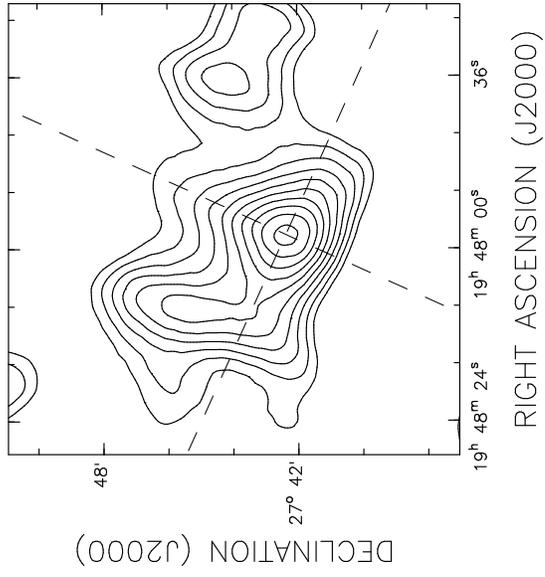
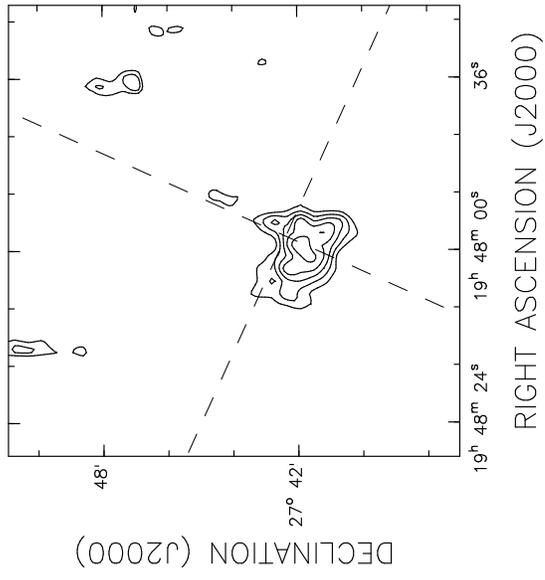
\end{document}